\documentclass[pra,twocolumn,superscriptaddress,showpacs,amsmath,amssymb]{revtex4}

\usepackage[dvips]{graphicx}

\newcommand{\be}{\begin{equation}}
\newcommand{\ee}{\end{equation}}
\newcommand{\ve}[1]{\mbox{\boldmath $#1$}}

\begin{document}

\title{Tkachenko modes in a superfluid Fermi gas at unitarity}
\author{Gentaro Watanabe}
\affiliation{Dipartimento di Fisica, Universit\`a di Trento and CNR-INFM R\&D
Center on Bose-Einstein Condensation, Via Sommarive 14, I-38050 Povo, Trento,
Italy}
\affiliation{The Institute of Chemical and Physical Research (RIKEN), 
2-1 Hirosawa, Wako, Saitama 351-0198, Japan}
\author{Marco Cozzini}
\affiliation{Dipartimento di Fisica, Politecnico di Torino, Corso Duca degli
Abruzzi 24, I-10129 Torino, Italy}
\affiliation{Quantum Information Group, Institute for Scientific Interchange
(ISI), Viale Settimio Severo 65, I-10133 Torino, Italy}
\author{Sandro Stringari}
\affiliation{Dipartimento di Fisica, Universit\`a di Trento and CNR-INFM R\&D
Center on Bose-Einstein Condensation, Via Sommarive 14, I-38050 Povo, Trento,
Italy}

\date{\today}

\begin{abstract}

We calculate the frequencies of the Tkachenko oscillations of a vortex lattice
in a harmonically trapped superfluid Fermi gas. We use the elasto-hydrodynamic
theory by properly  accounting for the elastic constants, the Thomas-Fermi
density profile of the atomic cloud, and the boundary conditions. Thanks to the
Fermi pressure, which is responsible for larger cloud radii with respect to the
case of dilute Bose-Einstein condensed gases, large vortex lattices are
achievable in the unitary limit of infinite scattering length, even at
relatively small angular velocities. 
This opens the possibility of experimentally observing vortex
oscillations in the regime where the dispersion relation approaches the
Tkachenko law for incompressible fluids and the mode frequency is almost
comparable to the trapping frequencies.

\end{abstract}


\maketitle

Two component Fermi gases are very versatile systems, offering the possibility
of exploring both fermionic and bosonic superfluidity. By exploiting Feshbach
resonances to tune the interaction strength between different spin species, it
is possible to realize ultracold gases of fermionic atoms and  Bose-Einstein
condensates (BECs) of dimers as well (for a recent review see
Ref.~\cite{giorgini07}). Moreover, at resonance, where the interspecies
scattering length diverges, fermionic atoms enter a new strongly correlated
regime, the so-called unitary limit, in which the standard
Bardeen-Cooper-Schrieffer (BCS) theory is not applicable. Here fermionic
superfluidity is significantly enhanced, due to the stronger role played by the
interactions.  Vortex lattices have been already observed in this novel phase
of matter \cite{zwierlein05}. The observation of vortices in Fermi gases
represents a particularly important achievement, since the unambiguous
detection of superfluidity in these systems is less straightforward than in the
corresponding case of Bose-Einstein condensates.

In this paper we study the Tkachenko modes of the vortex lattice in a
harmonically trapped Fermi gas at unitarity. These oscillations, originally
studied in Ref.~\cite{tkachenko66} for incompressible superfluids, correspond to
shear distortions of the lattice planes and carry much lower energies than usual
hydrodynamic modes. The investigation of Tkachenko modes in Bose-Einstein
condensates  has been largely pursued on both the theoretical 
\cite{anglin,baym03,cozzini04,baksmaty04,mizushima04,sonin05} and experimental
\cite{JILAtka,JILALLL} sides. An appealing reason to study vortex modes in  Fermi
gases at unitarity is given by  the larger number of vortices $N_v$ achievable
for a given rotation rate. As we shall discuss in the final part of this paper,
unitary fermions of $^6$Li yield a significant enhancement (almost an order of
magnitude) in the vortex number compared to bosonic atoms of $^{87}$Rb. This is
due to the large expansion of the cloud radius produced by the Fermi pressure.
This effect provides the possibility of studying Tkachenko modes in a more
systematic way than in the bosonic case, including  an easier detection of  modes
with more than one radial node  and the observation of Tkachenko modes at
relatively small angular velocity $\Omega$, where the incompressible regime of
the dispersion relation is exploited.

A first insight into the problem can be gained from the frequency spectrum of the
Tkachenko oscillations of an infinite vortex lattice \cite{sonin87}
\be\label{eq:homo}
\omega^2 = \frac{c_T^2c_s^2k^4}{4\Omega^2+c_s^2k^2} \ ,
\ee
where $k$ is the wave vector, $c_s$ is the usual sound velocity,
$c_T=\sqrt{\kappa\Omega/8\pi}$ is the Tkachenko velocity, and the limit
$c_s\gg{}c_T$ is assumed. Here $\kappa=h/M$ is the quantum of circulation of a
single vortex line, where $M=m$ for bosons and $M=2m$ for fermions, $m$ being the
mass of the particles. The incompressible limit of Eq.~(\ref{eq:homo}) takes
place for $c_sk\gg\Omega$ and corresponds to the original Tkachenko dispersion
law $\omega = c_Tk$. In the opposite limit $c_sk\ll\Omega$ one finds the
quadratic dispersion relation $\omega = c_Tc_sk^2/2\Omega$, characterizing the
compressible limit of the spectrum. In a non-uniform configuration the actual
values of $k$ are inversely proportional to the radius of the cloud and are fixed
by the proper procedures of discretization that will be discussed later. In the
experiments with BECs it is difficult to realize large vortex lattices at small
angular velocities and the condition $c_sk\gg\Omega$ is hence hard to fulfill.
Vice versa, we will show that in the case of fermions at unitarity the crossover
from incompressible to compressible behavior can be more easily investigated. 

In our calculation, we adapt the two-dimensional (2D) elasto-hydrodynamic
treatment developed by Sonin \cite{sonin05} to the fermionic case. We consider
two distinct density profiles in the Thomas-Fermi (TF) approximation: (i) the one
corresponding to the cylinder geometry as considered in Ref.~\cite{sonin05} and
(ii) the 2D column density obtained by integrating a 3D TF cloud along the axial
direction (hereafter also called pancake geometry). The cylinder geometry is an
idealized configuration whose interest lies in the exact decoupling between the
radial and axial dynamics. However, experimental situations are expected to be
better described by the pancake geometry, which takes into account the 3D
features of the inhomogeneous profile. While in the full 3D geometry the
decoupling of dynamics is not exact, the axial motion is expected to be
negligible for Tkachenko modes as long as vortices are straight. The use of the
column density is hence expected to be a valuable approximation for pancake shape
clouds which do not exhibit significant vortex bending.  

One of the major differences between Bose and Fermi gases lies in the equation of
state, being dominated in the Fermi case by the quantum pressure effect. In the
TF approximation, the equation of state can be expressed as
$\mu\propto{n}^\gamma$ in both cases, but with different values of the polytropic
index $\gamma$.  Here $\mu$ is the chemical potential and ${n}$ is the number
density of particles. In 3D, $\gamma=1$ for bosons and $\gamma=2/3$ for  fermions
at unitarity \cite{note_pure2d}, yielding a different TF density profile in the
two cases. For a vortex lattice rotating at an angular velocity $\Omega$ in a
harmonic trap with radial and axial frequencies $\omega_\perp$ and $\omega_z$,
the coarse-grained equilibrium density profile is given by $n(\ve{r})\propto
[\mu_0 - \tilde{V}(\ve{r})]^{1/\gamma}$, where $\mu_0$ is the chemical potential
for the trapped case, $\tilde{V}(\ve{r})\equiv
m(\tilde{\omega}_{\perp}^2r_{\perp}^2+\omega_z^2z^2)/2$  is the effective
potential taking into account the centrifugal force,
$\tilde{\omega}_{\perp}^2\equiv\omega_{\perp}^2-\Omega^2$, and
$r_{\perp}^2\equiv{}x^2+y^2$.

For a TF configuration with pancake geometry we may reduce the problem to two
dimensions by integrating out the $z$ coordinate; i.e., we employ the column
density $n_{\rm 2D}(r_{\perp})=\int\mathrm{d}z\,{n}(\ve{r})$ as an effective 2D
density profile. One hence finds $n_{\mathrm{2D}}(r_{\perp})\propto
(\mu_0-m\tilde{\omega}_{\perp}^2r_{\perp}^2/2)^{1/\gamma_{\rm eff}}$ with
$\gamma_{\rm eff}\equiv2\gamma/(2+\gamma)$ corresponding to the equation of
state $\mu \propto n^{\gamma_{\rm eff}}$  in the uniform 2D configuration. Note
that the effective 2D polytropic index $\gamma_{\rm eff}$ always differs from
the one calculated in the cylindrical geometry, where $\omega_z=0$ and
$\gamma_{\rm eff}=\gamma$. It is also worth pointing out that $\gamma_{\rm
eff}$ has the same value for fermionic systems in the cylindrical geometry and
for bosonic systems in the pancake geometry (see Table \ref{table_gammaeff}).
In the remainder of this paper, we treat the problem only in two dimensions
and  drop the subscript ``$\perp$'' of $r_{\perp}$ and the subscript ``2D'' of 
$n_{\rm 2D}$.

Macroscopic manifestations of superfluidity at zero temperature can be studied
by resorting to the superfluid hydrodynamic equations.  As long as we study the
dynamics of vortices on length scales much larger than the intervortex
distance, we can employ the  elasto-hydrodynamic theory
\cite{sonin76,baym83,sonin87}. Here the microscopic density and velocity fields
are substituted by averaged quantities and the restoring force of the lattice
is taken into account by adding an  elastic energy term to the usual superfluid
hydrodynamic energy functional \cite{note_bs}.

The elastic energy for a triangular lattice is given by
$E_{\textrm{el}}=\int\mathrm{d}{\ve{r}}\,\mathcal{E}_{\textrm{el}}$, where the
energy density $\mathcal{E}_{\textrm{el}}$ in the rotating frame is
$\mathcal{E}_{\mathrm{el}}=2C_1(\ve{\nabla}\cdot\ve{\epsilon})^2+C_2
[\left(\partial\epsilon_x/\partial x - \partial\epsilon_y/\partial y \right)^2+
\left(\partial\epsilon_x/\partial y+ \partial\epsilon_y/\partial x \right)^2]$.
Here $\ve\epsilon$ is the vortex displacement field, and the coefficients $C_1$
and $C_2$ are the compressional and shear modulus respectively \cite{baym83},
corresponding to second derivatives of the energy density with respect to
lattice distortions. They have to be calculated from the (microscopic) energy
functional evaluated in the rotating frame \cite{baym83}.

In the Fermi case at unitarity the core size $\xi$ of vortices is of the order
of the interparticle distance. Consequently, unless one works extremely close
to the centrifugal limit \cite{antezza07}, $\xi$ is much smaller than the
intervortex distance, which is of the order of the Wigner-Seitz radius of the
vortex lattice cell, namely,
$l_\Omega=\sqrt{\kappa/2\pi\Omega}=\sqrt{\hbar/M\Omega}$. This relation
corresponds to the usual vortex density $n_v=1/\pi{}l_\Omega^2=2\Omega/\kappa$
obtained from the condition of quantized circulation. The elastic coefficients
can then be calculated in the small core limit $\xi\ll{}l_\Omega$, equivalent
to the Thomas-Fermi condition $\mu_0\gg\hbar\Omega$ \cite{comprcond}. In this
regime, the well-established result $C_2=-C_1={n}m\kappa\Omega/16\pi$ holds
\cite{tkc2,baym83,sonin87}. Hence, for bosons $C_2=-C_1={n}\hbar\Omega/8$,
while for fermions $C_2=-C_1={n}\hbar\Omega/16$, i.e., in the latter case ${n}$
is replaced by the density of pairs, ${n}/2$. This result can be understood by
observing that fermionic pairs play the same role as bosonic molecules from the
point of view of superfluidity.

\begin{table}
    \caption{Effective polytropic index $\gamma_{\rm eff}$ in 2D
     for the bosonic and fermionic cases.  For the cylindrical
     geometry, $\gamma_{\rm eff}=\gamma$, and for the pancake
	 geometry, $\gamma_{\rm eff}=2\gamma/(2+\gamma)$. The values of the
parameter $\alpha$ characterizing the dispersion law in the incompressible
limit (see text) are also reported.}
\label{table_gammaeff}
\tabcolsep=5mm
\begin{tabular}{c|c|c}
\hline\hline
 geometry\quad & bosons & fermions \\
\hline
 cylinder &  1 ($\alpha=5.43$)   &  2/3 ($\alpha=5.59$)  \\
 pancake  &  2/3 ($\alpha=5.59$) &  1/2 ($\alpha=5.75$)  \\
\hline\hline
\end{tabular}
\end{table}

We are finally ready to write the linearized elasto-hydrodynamic equations. In the rotating frame they take the form
\begin{eqnarray}
\label{eq:elHD1}
\frac{\partial}{\partial t}\,\delta{n}+
\ve\nabla\cdot({n}_0\,\delta\ve{v}) & = & 0 \ , \\
\label{eq:elHD2}
\frac{\partial}{\partial t}\,\delta\ve{v}+
2\,\ve\Omega\wedge\delta\ve{v}+\ve\nabla\,\frac{\delta\mu}{m}
-\frac{\ve{F_{\textrm{el}}}}{m{n}_0} & = & 0 \ , \\
\label{eq:elHD3}
\frac{\partial}{\partial t}\,\delta\ve{v}+
2\,\ve\Omega\wedge\dot{\ve\epsilon}+
\ve\nabla\,\frac{\delta\mu}{m} & = & 0 \ ,
\end{eqnarray}
where $\dot{\ve\epsilon}=\partial\ve\epsilon/\partial t$
and the variation of the local chemical potential,
$\delta\mu$, can  be expressed in terms of the local
sound velocity as
$\delta\mu/m=(\partial\mu/\partial{n})
\delta{n}/m=(c_s^2/{n}_0)\delta{n}$.
By combining Eqs.~(\ref{eq:elHD2}) and (\ref{eq:elHD3}) one also finds
$2m{n}_0\ve\Omega\wedge(\delta\ve{v}-\dot{\ve\epsilon}) =
\ve{F_{\textrm{el}}}$.
Here ${n}_0\equiv{n}_0(\ve{r})$ is the equilibrium density, while $\delta{n}$
and $\delta\ve v$ are the density and velocity perturbations. The elastic force
is given by ${(F_{\textrm{el}})}_j=\partial\sigma_{jk}/\partial{x_k}$, where
$\sigma_{jk}=\delta{E}_{\textrm{el}}/\delta{u_{jk}}=
\partial\mathcal{E}_{\textrm{el}}/\partial{u_{jk}}$ is the stress tensor
defined in terms of the strain tensor
$u_{jk}=(\partial\epsilon_j/\partial{x_k}+\partial\epsilon_k/\partial{x_j})/2$
\cite{baym83}. Note that, while in the case of BECs for the cylindrical
geometry the ratio $c_s^2/{n}_0$ is constant and commutes with the gradient in
Eqs.~(\ref{eq:elHD2}) and (\ref{eq:elHD3}), this is not the case for
$\gamma_{\rm eff}\ne 1$.

We are interested in the axisymmetric Tkachenko modes. In polar coordinates all
the physical quantities are then independent of the azimuthal angle $\phi$ and
we can reduce to ordinary differential equations with respect to $r$.
For large vortex numbers one can approximate
$\dot\epsilon_\phi\simeq\delta{v}_\phi$ (see Ref.~\cite{sonin05}) and the above
linearized equations take the simplified form
\begin{eqnarray}
2i\omega\Omega\,\delta{v}_r & = & -\omega^2 \delta{v}_\phi
-\frac{c_T^2}{{n}_0}\frac{1}{r^2}\frac{\partial}{\partial r}
\left[{n}_0r^3\frac{\partial}{\partial r}
\left(\frac{\delta{v}_\phi}{r}\right)\right] \ , \quad \label{elasteq}\\
2i\omega\Omega\,\delta{v}_\phi & = &
\frac{\partial}{\partial r}\left[\frac{c_s^2}{{n}_0}
\frac{1}{r}\frac{\partial}{\partial r}
\left({n}_0r\delta{v}_r\right)\right] \ ,\label{eulereq}
\end{eqnarray}
where $\delta{v}_r$ and $\delta{v}_\phi$ are the radial and azimuthal components
of $\delta\ve{v}$, and $c_T=\sqrt{\kappa\Omega/8\pi}=l_\Omega\Omega/2$.

We solve Eqs.~(\ref{elasteq}) and (\ref{eulereq}) with proper boundary
conditions at the cloud center and at the cloud radius
$R=\sqrt{2\mu/m\tilde\omega_\perp^2}$ using the shooting method. At $r=0$ the
velocity must vanish: $\delta{v}_r(0)=\delta{v}_\phi(0)=0$. Following the same
arguments of Ref.~\cite{sonin05}, we obtain at $r=R$
\begin{eqnarray}
\frac{\partial}{\partial{r}}\delta{v}_\phi(R)-\frac{\delta{v}_\phi(R)}{R} & = & 0 \ , \label{bc_vt}\\
\frac{\partial}{\partial{r}}\delta{v}_r(R)+\frac{\delta{v}_r(R)}{R} & = &
-\frac{\gamma_{\rm eff}}{\gamma_{\rm eff}+1}\frac{i\omega\Omega
R}{c_s^2(0)}\delta{v}_\phi(R) \ .\label{bc_vr}
\end{eqnarray}

\begin{figure}
\includegraphics[width=7cm]{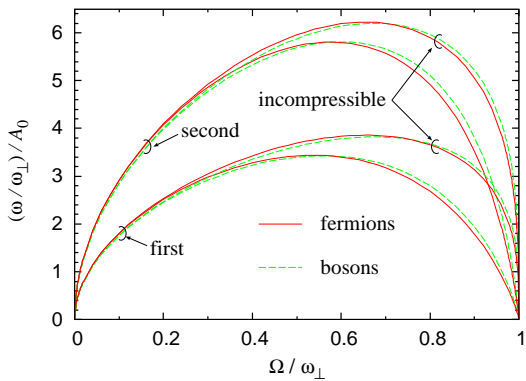}
\caption{\label{fig_q}(Color online)\quad Plot of the frequency $\omega$ of
  the first and the second axisymmetric Tkachenko modes as a function of the angular velocity.
  The green dashed line corresponds to pancake bosons ($\gamma_{\mathrm{eff}}=2/3$), while the red solid line to pancake fermions ($\gamma_{\mathrm{eff}}=1/2$). For each mode the incompressible limit $\omega_{\mathrm{inc}}=\omega_0\alpha$ is shown for comparison.
  }
\end{figure}

In Fig.~\ref{fig_q} we present our predictions for the frequencies of the two
lowest Tkachenko modes at unitarity.  In the numerical calculation it is
natural to solve Eqs.~(\ref{elasteq}) and (\ref{eulereq}) for the dimensionless
ratio $\omega/\omega_0$, where $\omega_0=c_T/R$. Since the discretized values
of $k$ are proportional to $1/R$ the  frequency $\omega_0$ is basically the
analogue of the incompressible limit $c_Tk$ of Eq.~(\ref{eq:homo}). Comparison
with the homogeneous spectrum of Eq.~(\ref{eq:homo}) then shows that
$\omega/\omega_0$ is the analogue of  $c_sk/\sqrt{4\Omega^2+c_s^2k^2}$, i.e.,
of the factor responsible for compressibility effects. The dependence of
$\omega_0$ on the angular velocity can be expressed analytically. The radius of
the cloud is given by $R=R_0/(1-\tilde\Omega^2)^\nu$, where $R_0$ is the radius
of the non-rotating cloud, $\tilde\Omega=\Omega/\omega_\perp$, and
$\nu=[2(\gamma_{\mathrm{eff}}+1)]^{-1}$. Then,
$\omega_0=A_0\omega_\perp\sqrt{\tilde\Omega}(1-\tilde\Omega^2)^\nu$, where
$A_0=\sqrt{m/4M}a_\perp/R_0$ and $a_\perp=\sqrt{\hbar/m\omega_\perp}$. The full
dependence of $\omega$ on $\Omega$ is hence displayed by plotting the quantity
$\omega/A_0\omega_\perp$ which does not depend any longer on $R_0$, but only on
the ratio $\Omega/\omega_\perp$. In the figure we also report the predictions
for a dilute Bose-Einstein condensed gas in the same pancake configuration. The
differences are actually very small, showing that, when expressed in the units
of Fig.~\ref{fig_q}, the results for the Tkachenko frequencies do not depend in
an appreciable way on the actual form of the equation of state which instead
can significantly affect the value of  the TF radius and hence of $A_0$.

One can get a qualitative understanding of the dependence of $\omega$ on
$\Omega$ in the trapped case looking at  the homogeneous dispersion
relation~(\ref{eq:homo}). To this purpose, one has to evaluate $c_s$ and $k$ in
the finite size system. A simple estimate for the sound velocity is given by
its value at the center of the cloud, i.e.,
$c_s^2\sim\gamma_{\mathrm{eff}}\mu/m=
\gamma_{\mathrm{eff}}\tilde\omega_\perp^2{R}^2/2$. The effective wave vector can
instead be quantized proportionally to $1/{R}$, where the proportionality
coefficient can be extracted by comparing Eq.~(\ref{eq:homo}) with the
numerical calculation. For a general value of  $\Omega$, the latter result
would depend on the chosen estimate for $c_s$. However, in the $\Omega\to0$
limit Eq.~(\ref{eq:homo}) becomes independent of $c_s$ and one can unambiguously
define $k=\alpha/R$ with $\alpha=\lim_{\Omega\to0}\omega/\omega_0$. The
quantized value of $k$ can then be used to estimate the full $\Omega$ 
dependence using Eq.(\ref{eq:homo}). This procedure was first exploited by Baym
\cite{baym03} in the case of BECs and yields a rather accurate estimate of the
whole dispersion law \cite{sound}. We find that the value of $\alpha$ exhibits
only a weak dependence on $\gamma$ and on the geometry employed (see Table 
\ref{table_gammaeff}). For the lowest Tkachenko mode we find  the value
$\alpha=5.75$ for fermions in the pancake geometry  to be compared with the
result $\alpha=5.59$ holding for bosons in the same geometry (or for fermions
in the cylindrical geometry) \cite{alpha}. Note also that the incompressible
limit $\omega_{\mathrm{inc}}=\omega_0\alpha$ of the dispersion relation
reproduces the full dispersion with high accuracy up to values $\tilde\Omega
=0.4\sim0.5$ of the angular velocity (see Fig.~\ref{fig_q}). 

Let us now discuss the dependence of the number of vortices $N_v$ on the experimental parameters. This is important in order to determine the
physical regimes of $\Omega/\omega_\perp$ achievable in practice.
The number of vortices in the TF approximation is $N_v=n_v\pi{}{R}^2$.
For a harmonically trapped Fermi gas at unitarity one has \cite{2D=3D}
$R=a_\perp[2\sqrt{1+\beta}
(3N\lambda)^{1/3}]^{1/2}(1-\tilde\Omega^2)^{-1/3}$, where $\beta\simeq-0.6$ is a universal dimensionless parameter accounting for the role of interactions 
\cite{qmc}, $N$ is the number of particles, and $\lambda=\omega_z/\omega_\perp$
is the trap aspect ratio. Thus,
$N_v(\mbox{fermions})=
4\sqrt{1+\beta}\,\tilde\Omega(1-\tilde\Omega^2)^{-2/3}(3N\lambda)^{1/3}$.
Using the corresponding expression for a Bose-Einstein condensed gas we find
\be
\frac{N_v(\mbox{fermions})}{N_v(\mbox{bosons})} =
\frac{4\sqrt{1+\beta}}{(1-\tilde\Omega^2)^{1/15}}
\frac{(3N\lambda)_f^{1/3}}{(15N\lambda{}a/a_\perp)_b^{2/5}} \ ,
\ee
where the subscripts refer to bosons and fermions and $a$ is the $s$-wave scattering length. For
$(N\lambda)_f=(N\lambda)_b$ the above ratio becomes
$4\sqrt{1+\beta}\ [3N\lambda(1-\tilde\Omega^2)]^{-1/15}(5a/a_\perp)_b^{-2/5}$ and
the small exponent $-1/15$ implies that the result is practically insensitive 
to the value of $\tilde\Omega$, $N$, and $\lambda$. On the other hand, in typical BECs one has
$a/a_\perp\ll1$, so that $N_v(\mbox{fermions})/N_v(\mbox{bosons})$ can be
significantly larger than 1 \cite{radius}.

In the JILA experiments on BECs one has $\lambda=0.63$, $a\simeq5\,$nm,
$a_\perp=3.74\,\mu$m, and $N\sim10^6$ \cite{JILAtka}. This yields
$N_v(^{87}\mathrm{Rb})=43.7\tilde\Omega(1-\tilde\Omega^2)^{-3/5}$. In the MIT
experiments, at unitarity one has $\lambda=0.40$ and
$N\sim10^6$ \cite{zwierlein05}. Then $N_v(^6\mathrm{Li})/N_v(^{87}\mathrm{Rb})\simeq6.1(1-\tilde\Omega^2)^{-1/15}$, which corresponds
to the significant gain in the number of vortices mentioned 
in the introduction \cite{note_3body}.

The increase of $N_v$ at unitarity allows to measure Tkachenko modes at
relatively small angular velocities. For example, a fermionic cloud with the
above parameters  easily contains more than $100$ vortices at
$\tilde\Omega=0.4$, deeply in the incompressible region of the spectrum.  The
maximum of the frequency of the lowest Tkachenko mode  takes place at
$\tilde\Omega\simeq0.55$ and corresponds to $\omega/\omega_\perp\simeq0.13$ for
the same parameters. This value is quite larger than the highest frequency
observed in $^{87}$Rb experiments, namely, $\omega/\omega_\perp=0.023$ at
$\tilde\Omega=0.84$ \cite{JILAtka}. This provides promising perspectives for
precision measurements of Tkachenko modes \cite{note_imaging}, for a direct
determination of the quantum of circulation in Fermi superfluids, and for the
general investigation of the incompressible regime of the dispersion relation,
whose quantitative analysis in experiments is a long standing question in the
field of superfluid systems.

\acknowledgments

Interesting discussions with M.~W. Zwierlein are acknowledged.
This work is supported by the Ministero
dell'Istruzione, dell'Universit\`{a} e della Ricerca (MIUR).


\begin{thebibliography}{99}
\bibitem{note_pure2d} In the pure 2D situation, where $\hbar\omega_z\gg \mu$
($\omega_z$ is the axial trap frequency) and the 2D equation of state holds,
$\gamma=1$ for both Bose and Fermi gases.
\bibitem{note_bs}
In the case of Fermi superfluids, there exist fermionic bound states localized
in the vortex cores \cite{caroli64} whose energy, at unitarity, 
is of order of the Fermi energy \cite{sensarma06}.  Since the Tkachenko modes 
have much smaller energy, the coupling between these two kinds of excitations 
can be safely ignored.
\bibitem{comprcond}
This condition should not be confused with the condition
$\Omega^2/c_s^2k^2\sim\hbar\Omega N_v/\mu_0\ll 1$ characterizing the
incompressible limit of the Tkachenko dispersion law.
\bibitem{sound}
The estimate becomes even more accurate if one employs the sound velocity
averaged over the whole cloud \cite{baym03}.
\bibitem{alpha}
For $\gamma_{\mathrm{eff}}=1$ one recovers the value $\alpha=5.43$
corresponding to the incompressible 2D result of Ref.~\cite{anglin} (see
\cite{note_pure2d}), also confirmed in Refs.~\cite{cozzini04,sonin05}. We also
note that the value $\alpha=5.59$ for the bosonic column density, corresponding
to a 3D pancake geometry, agrees with the value extracted from the 3D sum rule
calculation of Ref.~\cite{cozzini04}.
\bibitem{radius}
Since the Tkachenko frequency scales like
$\omega\propto1/R\propto1/\sqrt{N_v}$, the larger value of $N_v$ quenches the
Tkachenko frequency at fixed $\Omega$. For example, for the experimental values
considered in the text, the value of $A_0$ for bosons is about a factor $2.6$
larger than for fermions. However, as we shall see later,  this quenching can
be largely compensated by choosing relatively small rotation rates.
\bibitem{note_3body}
In principle, one can increase $N_v$ in BECs by employing larger values of $a$
and/or $N$. However, three-body recombination losses become crucial for large
$a$ and/or $N$, which makes difficult to obtain very large vortex lattices 
unless one works close to the centrifugal limit.
\bibitem{note_imaging}
Due to the extremely small vortex core size, the imaging of vortices in atomic
Fermi gases at unitarity is actually a non-trivial issue. This difficulty can
be overcome by ramping the scattering length to small and positive values just
before the expansion.
\end{thebibliography}
\end{document}